# Reflection statistics of weakly disordered optical medium when its mean refractive index is different from an outside medium

Prabhakar Pradhan[1*], Daniel John Park[2], Ilker Capoglu[2], Hariharan Subramanian[2], Dhwanil Damania[2], Lusik Cherkezyan[2], Allen Taflove[3], Vadim Backman[2]

[1]*Department of Physics, BioNanoPhotonics Laboratory, University of Memphis, Memphis, TN 38152*
[2]*Biomedical Engineering Department, Northwestern University, Evanston, IL 60208*
[3]*Electrical Engineering and Computer Science Department, Northwestern University, Evanston, IL 60208*
[*]*ppradhan@memphis.edu*

**Abstract:** Based on the difference between an optical sample's mean background refractive index $n_0$ and an outside medium $n_{out}$ ($\neq n_0$), we study the reflection statistics of a one-dimensional weakly disordered optical medium with refractive index $n(x) = n_0 + dn(x)$. Considering $dn(x)$ as color noise with the exponential spatial correlation decay length $l_c$ and $k$ as the incident wave vector, our results show that for the small correlation length limit, i.e. $kl_c<1$, the average value of reflectance $r$ follows a form that is similar to that of the matched refractive-index case $n_0 = n_{out}$, i.e., $<r(dn, l_c)> \propto <dn^2> l_c$. However, the standard deviation of $r$ is proven to be $\sigma[r(dn,l_c)] \propto (<dn^2>l_c)^{1/2}$, which is different from the matched case. Applications to light scattering from layered media and biological cells are discussed.

©2015 Optical Society of America

**OCIS codes:** (290.0290) Scattering; (290.1350) Backscattering; (290.5825) Scattering theory

## 1. Introduction

The statistical transport properties of one-dimensional (1D) mesoscopic disordered optical and electronic media are now well studied [1-4, 16,17]. The Schrödinger equation and Maxwell's wave equation are similar in the sense that they can be projected to the Helmholtz equation; therefore, the formalisms are the same for corresponding scalar waves in both cases [5-8]. After the Landauer formalism showed that the reflection coefficient is related to the resistance/conductance of the sample, the outer scattering components such as the reflection and transmission coefficients became important for the study of localization and conductance fluctuations in the electronics case [5, 6]. Similarly, extending the ideas from the electronic systems, in previous studies of light scattering and localization properties of different optical disordered media the fluctuation part of the refractive index is primarily considered while the sample's mean refractive index is the same as the outside medium [5, 7-9]. The results show that both the average reflection and the fluctuations have the same form for the mesoscopic optical sample. However, the mismatch of the refractive index between the sample and the outside medium and its effect upon the reflection statistics remains poorly understood.

In this paper, we study reflection statistics in the context of the synergistic effects between refractive index mismatch values and the fluctuation of the refractive index. For a biological medium, for example a biological cell, the spatial fluctuation of the refractive index is quite small (~.01) and hidden within a higher uniform refractive index (~1.38). Enhancement of the backscattering signals from the weakly fluctuating refractive index, as mediated by the refractive index mismatch, is also addressed. Finally, applications of the method for light scattering from biological cells are discussed in terms of the enhancement of the signal from refractive index nano-fluctuations for potential application to cancer detection.

`

## 2. Reflection statistics of matched disordered media

Consider a 1D sample of length $L$ with refractive index inside the sample $n(x) = n_0 + dn(x)$ ($0 < x < L$), where the average refractive index of the samples is $n_0 = <n(x)>$, $dn(x)$ is the fluctuation part of the refractive index with $<dn(x)> = 0$, and $n_{out}$ is the refractive index of the outside medium as shown schematically in Fig. 1. The 'matched' case can be defined as *equality* between the average refractive index of the sample and the outside medium, i.e., $n_0 = n_{out}$, whereas the 'mismatched' case can be defined as $n_0 \neq n_{out}$. Since we are interested in reflection statistics, let us define $R(L)$ as the complex reflection amplitude of a sample of length $L$ which is illuminated by a plane wave of wave vector $k$. Then, the mean and standard deviation of reflectance ($r = RR^*$) are the primary concerns of this work. For example, we will prove below that the mean reflectance for the mismatched case, $<r_{mismatched}>$, can be written in terms of the reflectance of a slab ($n_0 \neq n_{out}$) and the mean reflectance of the matched case, $r_{slab}$ and $<r_{matched}>$, as:

$$<r_{mismatched}> = r_{slab} + <r_{matched}> \times F(k, n_0, L, r_{slab}). \qquad (1)$$

In the literature, reflection statistics from disordered optical media are primarily studied assuming the matched case, and the mismatched case, as defined above, is not well studied. Therefore, we first briefly review the results of a one-dimensional matched case ($r_{matched}$) before describing the results of the mismatched case ($r_{mismatched}$). The statistics of spatial random refractive index fluctuation, $dn(x)$, generally represented by color noise, i.e., $<dn(x)> = 0$ and $<dn(x) \cdot dn(x')> = <dn^2> \exp(|x-x'|/l_c)$, where $l_c$ is the spatial correlation decay length of the spatial refractive index fluctuation $dn(x)$. Then, using the Fokker-Planck approach, the above $<r_{matched}>$ can be solved analytically in the weakly disordered limit (i.e., $dn << n_0$) [5, 7]. The mean value of reflection and its standard deviation both have the same value, $<r_{matched}> = \sigma(r_{matched}) = L/\xi$, where the inverse of the localization length has the form $\xi^{-1} = 2k^2 <dn^2> l_c/[1+(2kl_c)^2]$. This is true for a weakly disordered sample where $\xi > L$, which can also be called a Born approximation to the weakly disordered part of the refractive index.

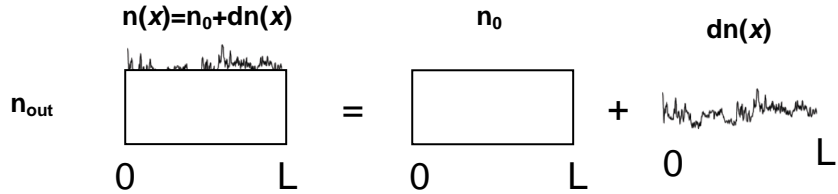

Fig. 1: Schematic of a mismatched case where outside refractive index is $n_{out}$ and sample refractive index is $n(x) = n_0 + dn(x)$, with $n_0$ as the average refractive index of the sample and $dn(x)$ as the spatial refractive index fluctuation of the sample $<dn(x)> = 0$. For matched case, $n_0 = n_{out}$.

## 3. Reflection statistics of mismatched disordered media

However, index mismatched weakly disordered samples are quite common for optical scattering experiments. For example, biological cells and tissues have refractive indices $n_0 \sim 1.3 - 1.5$ and $dn \sim 0.01 - 0.1$ with the outside air medium $n_{out} = 1$. In the case of weak refractive index fluctuations, the backscattering light transport properties of such biological cells can be decomposed into a multiple-transport 1D channel or a quasi-1D parallel multichannel problem [10]. It was recently shown that quasi-1D multichannel backscattering would provide sensitivity to changes in the nanoscale signal relative to a three-dimensional (3D) bulk for weakly disordered media such as biological cells. Furthermore, the approach has been proven to be useful for early pre-cancer screening by detecting changes in

`

the nanoscale refractive index fluctuations of cells related to the progress of carcinogenesis in different types of cancers [11-14].

To derive the form of Eq. (1), we start from a stochastic Langevin equation (here, stochasticity enters into the equation through the *dn(x)* term) for the index mismatched case ($n_{out} \neq n_0$) which gives the reflection amplitude $R_t$. For simplicity, we will consider that the sample is kept in air, i.e., $n_{out} = 1$ and $n_0 > 1$. Substituting these terms ($n_{out}$, $n_0$ and *dn*) with color noise, the Langevin equation for the mismatched case can be derived following the invariant imbedding approach [5]:

$$\frac{dR_t(L)}{dL} = 2ikR_t(L) + \frac{ik}{2}[(n_0^2 - 1) + 2n_0 dn(L)] \times [1 + R_t(L)]^2. \qquad (2)$$

The complex total reflection (amplitude $R_t(L)$ from a weakly disordered medium can be considered as a combination of: (i) a deterministic sinusoidal oscillation component $R_{slab}$ based on the pure background of a thin-film slab of length *L* without any stochastic *dn(x)* terms, and (ii) a *R* component that contains *dn(x)* terms. Therefore, $R_t = R_{slab} + R$. Each term can then be easily derived from Eq. (2) as follows:

$$R_t(L) = R_{slab}(L) + R(L), \qquad (3a)$$

$$\frac{dR_{slab}(L)}{dL} = 2ikR_{slab} + \frac{ik}{2}(n_0^2 - 1) \times [1 + R_{slab}]^2, \qquad (3b)$$

$$\frac{dR(L)}{dL} = 2ikR + \frac{ik}{2}(2n_0 dn(L)) \times [1 + R_{slab} + R]^2$$
$$+ \frac{i}{2}k(n_0^2 - 1) \times [2R(1 + R_{slab}) + R^2]. \qquad (3c)$$

In Eqs. (3a-c), the perturbative contribution by the stochastic term *dn(x)* has many cross-terms between $R_{slab}$ and *R*. We will assume that *R* is in the first order in *dn(x)*. By performing a phase transformation as below in Eqs. (3b-c), we can further simplify and assimilate the $R_{slab}$ - *R* cross-terms in the equation. For this, we introduce new variables, $Q(L)$ and $\alpha(L)$, which are derived from $R(L)$ by a phase transformation as follows:

$$R_{slab}(L) = Q_{slab}(L) \cdot e^{2ik\alpha(L)}, \qquad (4a)$$

$$R(L) = Q(L) \cdot e^{2ik\alpha(L)}. \qquad (4b)$$

This yields a new set of simplified equations for $Q(L)$ and $\alpha(L)$ which further simplifies to:

$$\frac{d\alpha(L)}{dL} = 1 + \frac{(n_0^2 - 1)}{2}(1 + R_{slab}), \qquad (5a)$$

$$\frac{dQ}{dL} = \frac{i}{2}k(2n_0 dn(L))e^{-2ik\alpha}[1 + R_{slab} + Qe^{2ik\alpha}]^2$$
$$+ \frac{i}{2}k(n_0^2 - 1)(Q)^2 e^{2ik\alpha}. \qquad (5b)$$

`

With the new representation above Eq. (5), the mean $<r_t>$ and the standard deviation $\sigma(r_t)$ of the reflectance for mismatched case $r_{\text{mismatched}} = r_t \equiv R_t R_t^*$ can be derived. By using Eq. (3a) and performing a realization average, so we obtain:

$$<r_t(L)> = <R_t R_t^*> = <(Q_{slab} + Q) \times (Q_{slab} + Q)^*>,$$
$$= r_{slab} + Q_{slab}^* <Q> + c.c. + <|Q|^2>, \qquad (6)$$

where $r_t = |R_t|^2$ and $r_{slab} = |R_{slab}|^2 = |Q_{slab}|^2 = \dfrac{(n_0^2 - 1)^2 \sin^2(n_0 k L)}{4 n_0^2 + (n_0^2 - 1)^2 \sin^2(n_0 k L)}.$

In Eq. (6), we have separated the slab pure reflection/interference contribution, $r_{slab}$, (when $dn=0$) and the disorder contribution. To calculate (6) and the standard deviation (see below), we calculate the following terms in leading order of $dn$ as:

$$<|Q|^2>, <(Q)^2>, <Q>,$$

which can be written explicitly as:

$$<|Q|^2> = -\frac{i}{2} k n_0 \int_0^L dL'[e^{2ik\alpha^*}(1 + R_{slab}^*)^2 \times <2dnQ> - c.c.], \qquad (7a)$$

$$<(Q)^2> = i k n_0 \int_0^L dL' e^{-2ik\alpha}(1 + R_{slab})^2 \times <2dnQ>, \qquad (7b)$$

$$<Q> = ik \int_0^L dL'[n_0(1 + R_{slab}) <2dnQ> + \frac{1}{2}(n_0^2 - 1) e^{-2ik\alpha} <(Q)^2>]. \qquad (7c)$$

By using Ornestein-Uhlenbeck stochastic process and Novikov theorem [5], the averaging of the above Eqs. (7a-c) was performed. For example, the disorder average of the product term $<2dn(L)R>$ is:

$$<2dn(L)R> = \frac{i}{2} k n_0 (\frac{g}{2}) \times [1 - \frac{\partial}{\partial(L/l_c)}] e^{-2ik\alpha}(1 + R_{slab})^2$$
$$+ O(dn^2 \cdot l_c^3) + O(dn^4 \cdot l_c^2), \qquad (8)$$

where the value of $g$ and $\alpha$ can be expressed as:

$$g = 8 <dn^2> l_c \quad \text{(disorder strength)}, \qquad (9a)$$

$$\alpha = \int_0^L ik\, [2 + (n_0^2 - 1) \times (1 + R_{slab})]\, dL'. \qquad (9b)$$

`

Finally, by using Eqs. (7-9) and averaging over the ensemble space, we can write the average $<r_t>$ in terms of $<dn^2>$ and $l_c$ as:

$$<r_{mismatched}> = <r_t> = r_{slab} + <dn^2> l_c \cdot k^2 [Q^*_{slab}(n_0^2 I_1 + kn_0^2(n_0^2 - 1)I_2) + c.c. + n_0^2 I_3 + c.c.] + O(dn^4), \quad (10)$$

where we have defined $I_1$, $I_2$, and $I_3$ in the above deterministic equations as:

$$I_1 = -2\int_0^L dL'(1 + R_{slab}) \times [1 - \frac{\partial}{\partial(L'/l_c)}]e^{-2ik\alpha}(1 + R_{slab})^2, \quad (11a)$$

$$I_2 = -i\int_0^L dL' e^{-2ik\alpha(L')} \int_0^{L'} dL'' e^{-2ik\alpha(L'')}(1 + R_{slab}(L''))^2 \times$$
$$[1 - \frac{\partial}{\partial(L''/l_c)}]e^{-2ik\alpha(L'')}(1 + R_{slab}(L''))^2, \quad (11b)$$

$$I_3 = \int_0^L dL' e^{-2ik\alpha^*}(1 + R^*_{slab})^2 \times [1 - \frac{\partial}{\partial(L'/l_c)}]e^{-2ik\alpha}(1 + R_{slab})^2. \quad (11c)$$

Eq. (10) can now be rewritten in terms of a pure slab term (without $dn(x)$) and the fluctuation term averages. For weak disorder with short range correlation $2kl_c<1$, the quantities $I_1$, $I_2$, and $I_3$ are approximately independent of $l_c$. In this case, we obtain.

$$<r_t> = <r_{mismatched}> = \quad (12)$$
$$= r_{slab} + \frac{1}{2}<dn^2> l_c \cdot k^2 L \times F(k, n_0, L, r_{slab}) + O[dn^4 l_c^2]$$
$$= r_{slab} + \frac{L}{\xi} \times F(k, n_0, L).$$

Thus, we can write a relationship between the matched and mismatched cases from Eq. (12):

$$<r_{mismatched}> = <r_t> = r_{slab} + <r_{matched}> \times F(k, n_0, L). \quad (13a)$$

The mean-square fluctuation of $r$, $\sigma^2(r)$, can be evaluated as:

$$\sigma^2(r_{mismatched}) = <r^2_{mismatched}> - (<r_{mismatched}>)^2$$
$$= <(r_{slab} + 2\text{Re}(Q^*_{slab}Q) + |Q|^2)^2>$$
$$- (<r_{slab} + 2\text{Re}(Q^*_{slab}Q) + |Q|^2>)^2. \quad (13b)$$

`

The above equation can now be further modified by applying Eqs. (4) and (7):

$$\sigma^2(r_{mismatched}) = Q^*_{slab}<Q> + c.c. + 2r_{slab}<|Q|^2> + O(dn^4 l_c^2). \quad (14a)$$

Taking the average over the disorder, we obtain a deterministic expression of the above equation:

$$\sigma^2(r_{mismatched}) = <dn^2> l_c \cdot k^2 n_0^2 [Q^*_{slab} I_4 + c.c. + 2r_{slab} I_3 + c.c.] + O(dn^4 l_c^2). \quad (14b)$$

For the mismatched case, it should be noted that the mean-square fluctuation of $r$ has a barrier or mismatch-induced leading order: $<dn^2>.l_c$. In the matched case, the leading order is higher, that is, $<dn^2>^2.l_c^2$. Therefore, we obtain the expression for the standard deviation $\sigma(r)$ as follows:

$$\sigma(r_{mismatched}) = <dn^2>^{1/2} l_c^{1/2} \cdot k \ [Q^*_{slab} I_4 + c.c. + 2r_{slab} I_3 + c.c.]^{1/2} + O(dn^2), \quad (15)$$

where we have defined:

$$I_4 = -2 \int_0^L dL' e^{-2ik\alpha}(1+R_{slab})^2 [1 - \frac{\partial}{\partial(L'/l_c)}] e^{-2ik\alpha}(1+R_{slab})^2. \quad (16)$$

Eq. (15) can be further written to the leading order as (following [5]):

$$\sigma(r_{mismatched}) = <dn^2>^{1/2} l_c^{1/2} \cdot k \ L^{1/2} [G(n_0, L, k)]. \quad (17)$$

where $G$ is a function without the $<dn^2>$ term. It can be noted that in the matched case ($n_0=1$), the value of $G$ is 0 since $Q_{slab}$ and $r_{slab}$ have the multiplicative factor $(n_0^2 - 1)=0$ as $n_0=1$. In this case, in $\sigma(r)$ the first order term in $<dn^2>^{1/2}$ vanishes and the second order term $<dn^2>$ is the leading term, recovering the matched case. Here we emphasize again that the value of $\sigma(r)$ calculated here with the assumption that the reflection from the fluctuating part is less than the reflection amplitude from the slab.

Eqs. (10) and (15) are plotted in Fig. 2(a) and 2(a') respectively for a sample length $L = 2\ \mu$; similarly, and in Fig. 2(b) and (b') for $L = 5\mu$. The other parameters remain the same: wavelength = 500 nm and $l_c$ = 20 nm.

We also performed direct stochastic simulation of Eq. (2) and then performed realization averages numerically (shown by the dotted lines) in Fig. 2 (a,a',b,b'). It can be seen that the theoretical results and the corresponding numerical results agree well for all of the parameters, thereby validating our theoretical calculations. For the case $2kl_c > 1$, detailed calculations will be reported in a separate paper.

`

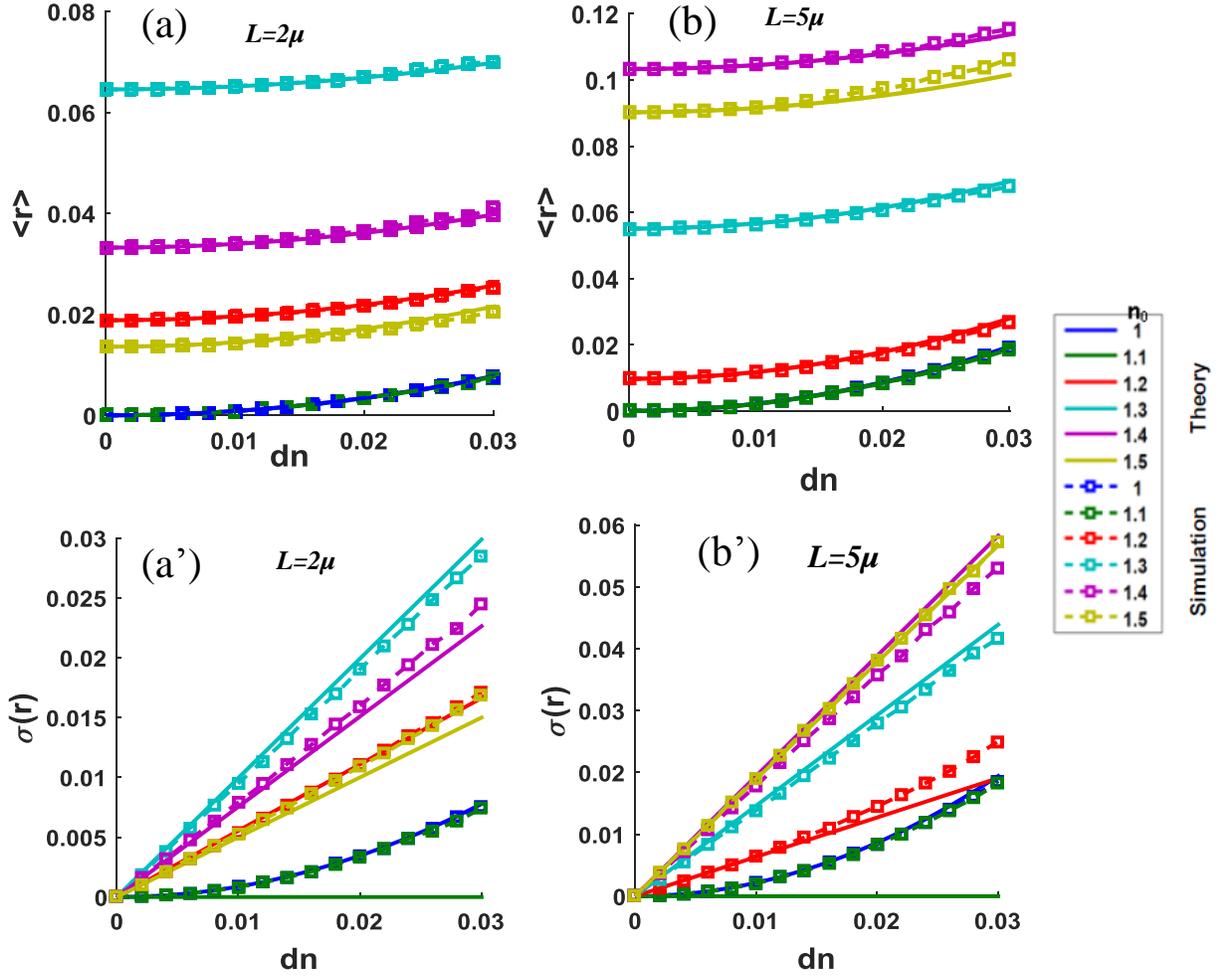

**Fig 2: (a)-(b)** Plots of analytical calculations (solid lines) of Eq. (12) with the pure numerical integration of Eq. (11) for $<r>$ and numerical simulations (squares) based on Eq. (2) with lengths: (a) $L = 2\mu$ and (b) $L = 5\mu$. (a')-(b') The plot of the analytical calculations (solid lines) of Eq. (15) and numerical simulations (squares) for $\sigma(r)$ based on Eq.(2) with lengths: (a') $L = 2\mu$ and (b') $L = 5\mu$. Refractive index of outside medium is taken as air $n_{out} = 1$. The mean refractive indexes of the samples are $n_0 = 1$ (blue, matched case), 1.1(green), 1.2(red), 1.3(cyan), 1.4(purple), and 1.5(yellow). Refractive index fluctuations were varied such that $dn = 0 - 0.03$. Correlation length of $dn$ spatial fluctuations $l_c = 20$ nm.

## 4. Conclusions and Discussions

In conclusion, our results show that the average reflectance, for the mismatched case with weak disorder and short range correlation, is linearly dependent on that of the matched case; the average reflectance $<r(dn, l_c)_{mismatched}>$ is proportional to $<dn^2>\cdot l_c$. However, the value of the standard deviation of the reflectance has a different form (Eq. (15)). This is because, as seen in Eq. (14(b)), the index mismatch parameter $(n_0^2 - 1)$ contributes (through $Q_{slab}=0$ and $r_{slab}=0$) and this changes the leading term of the mean-square fluctuations from $(<dn^2>.l_c)^2$ to $<dn^2>.l_c$. Therefore, the RMS fluctuations, or STD for the mismatched case, $\sigma(r(dn,l_c)_{mismatched})$ is proportional to $<dn^2>^{1/2}l_c^{1/2}$. The relative fluctuation $\sigma(r_{mismatched})/<r_{mismatched}>$ decreases ($<1$) with the increase of the mismatched parameter ($n_0^2 - 1$). However, the relative $\sigma(r_{mismatched})$ value for the mismatched case, compared to the matched case, increases.

The phenomena of the mismatched induced enhancement can be useful for enhancing a weak reflective signal from the weak refractive index fluctuations ($<dn^2>^{1/2}$) hidden in a strong uniform refractive index background ($n_0$), such as in biological cells. Our results show that the backscattering signal from the fluctuation part of the refractive index can be enhanced by increasing the mean background refractive index. This is due to the multiple reflections of the wave within the background allowing more/longer interactions between the wave within the stronger background boundaries and the imbedded refractive index fluctuations. Therefore, the developed method has potential applications for enhancing scattering from biological cells where the refractive index fluctuations ($<dn^2>^{1/2}$ varies from .001 to ~.05) are imbedded within the cell's background refractive index $n_0$~1.38-1.5, with mismatched refractive index ~.38-.5, with respect to the air medium index 1. Recently it has been shown that the nanoscale fluctuations in a cell increase with the progress of early carcinogenesis [11-12] due to the rearrangements of the cell's building blocks (DNA, RANA, Lipids *etc.*). Therefore, enhancing the nanoscale signal from biological cells by simply changing the background (i.e., slab) refractive index will increase cancer detection sensitivity. For example, one can easily enhance the reflection signal from the fluctuating part by simply increasing the background refractive index, such as by dipping or treating the cell in a nonreactive liquid having a higher refractive index than the cell (i.e., >1.38). This is important for practical applications, as just a simple cell treatment/preparation could enhance the diagnostic or sensitivity/specificity, and will have potential clinical applications in improving cancer detection. Our results further suggest the best way to obtain a stronger signal from the fluctuating parts in a cell is keeping the cell in a high-contrast situation. For example, a stronger signal from nano-fluctuations can be obtained when the cells are kept on a slide and exposed to air interface, producing a high-contrast situation, than when the cells are kept on a slide covered with a coverslip, producing a low-contrast situation.

Although we have considered here the one-dimensional back reflection case, the fact of signal enhancement from embedded fluctuation in a background will have similar physical reasons for higher dimensions. For a thinner sample (relative to the probing wavelength λ), such as cheek or buccal cells (width ~500nm), the back reflection can be treated as a bunch of quasi-one-dimensional parallel back-propagating interacting channels (area of a channel ~$\lambda^2$). It has been shown that the average value of the reflection from a channel in a multi-channel propagation decreases inversely with the number of channels [10]. Therefore, the solution of 1D channel provides us with an understanding of the signal enhancement and framework for more practical three-dimensional signal enhancement. Other than biological media, artificial layered media could be made easily [18] for active (with lasing or absorption) and passive cases, for 1D or quasi-1D systems for different applications. Therefore, we believe the present work will have many applications, varying from biological cells to varieties of layered media.

**Acknowledgements**

This work was partly supported by NIH grants (Nos. R01EB003682, R01CA128641 and U54CA143869) and NSF grant no CBET-0937987.

`

`